\documentclass[fleqn,usenatbib]{mnras}

\pdfoutput=1

\usepackage{lipsum}
\usepackage{latexsym}
\usepackage{amssymb}
\usepackage{graphicx}
\usepackage[utf8x]{inputenc}
\usepackage[figuresright]{rotating}
\usepackage{color,xspace}
\usepackage{longtable}
\usepackage{multicol}
\usepackage{supertabular}
\usepackage{float} 
\usepackage{pdfpages}
\usepackage[T1]{fontenc}
\usepackage{aecompl}

\usepackage{amsmath}	
\usepackage{textcomp}

\title[2D kinematic of the central region of NGC\,4501.]{2D kinematic study of the central region of NGC\,4501.}
\author[P. Repetto et al.]{P. Repetto$^{1}$\thanks{E-mail:prsatch6@gmail.com}, 
M. Fa\'{u}ndez-Abans$^{1}$, P. Freitas-Lemes$^{2, 4}$, I. Rodrigues$^{2}$, \newauthor M. de Oliveira-Abans$^{1,3}$
\thanks{Based on observations obtained at the Gemini Observatory acquired through the Gemini Science Archive.}\\
$^{1}$Laborat\'{o}rio Nacional de Astrof\'{i}sica, Rua Estados Unidos 154, 37504-364, Itajub\'{a}, MG, Brazil\\
$^{2}$Universidade do Vale do Para\'{i}ba, UNIVAP. Av. Shishima Hifumi, 2911, Urbanova CEP: 12244-000, S\~{a}o Jos\'{e} dos Campos, SP, Brazil\\
$^{3}$UNIFEI, Instituto de Engenharia de Produ\c{c}\~{a}o e Gest\~{a}o, Av. BPS 1303, Pinheirinho, 37 500-903, Itajub\'{a}, MG, Brazil\\
$^{4}$Departamento de Física, Instituto Tecnológico de Aeronáutica (ITA), São José dos Campos, 12228-900, SP, Brazil}

\pagerange{\pageref{firstpage}--\pageref{lastpage}} \pubyear{2015}

\begin{document}
\label{firstpage}
\date{Accepted xxxx XXXXXXXX 15. Received 2014 January 15; in original form 2014 January 15}
\maketitle

\begin{abstract}
GMOS-IFU observational data were used to study the detailed two dimensional gas kinematics and morphological structures within the $\sim\,500 \times 421$ pc$^2$ of the
active \mbox{Seyfert 2} galaxy NGC\,4501. We provide empirical evidences of possible outflowing material from the central zones of NGC\,4501 to the observer. In addition,  
we performed a spectral synthesis and diagnostic diagram analysis to determine respectively the dominant stellar population in the inner disc of this galaxy and to unveil
the actual nature of the central engine of NGC\,4501. The principal finding of this work is that the central regions of NGC\,4501 are dominated by non circular motions 
connected to probable outflows of matter from the nuclear regions of this galaxy. A predominant old stellar population inhabits the internal zones of NGC\,4501 excluding
the possibility of ongoing starburst activity in the central parsecs of this galaxy. The latter result is confirmed by the diagnostic diagram analysis that establishes a
preponderant active galactic nucleus character for NGC\,4501. These outcomes together provide a general description of the gas motion and the 
corresponding nuclear activity in the internal disc of NGC\,4501 in an attempt to elucidate the possible relation among the central activity and the induced kinematic 
properties of this nearby galaxy.
\end{abstract}


\section{Introduction}\label{sec:s1}

Nearby galaxies with active galactic nucleus (AGN) are good laboratories to study and understand how mass is transferred from galactic scales down to nuclear scales to feed a 
supermassive black hole. Theoretical models and simulations have suggested that non-axisymmetric potentials promote gas inflow towards the inner regions of 
an AGN \citep{Shlosman1990,Emsellem2003,Knapen2005,Englmaier2004}. Observations of the inner kiloparsec of AGNs revealed the existence of some morphological 
structures such as inner small-scale disks, nuclear bars and spiral-like arms \citep{Erwin1999, Pogge2002, Laine2003}. \citet{Martini2003} estimated that in 
more than half of AGNs the most common nuclear kiloparsec structures are dusty spirals. This supports the idea that those inner spiral-like arms could be part 
of the fuelling chain to the central engine, by transporting gas from the outside kiloparsec down to a few tens of parsecs 
\citep{Knapen2000, Emsellem2001, Maciejewski2002, Marconi2003, Crenshaw2003, Fathi2005, Fathi2006, Maciejewski2004}. \citet{Fathi2005} and 
\citet{Storchi-Bergmann2007} mapped the streaming motion of gas towards the nuclei of the active galaxies NGC\,1097 SB(s)b, and NGC\,6951 SAB(rs)bc. 
\citet{Prieto2005} discovered spiral structure within the inner \mbox{300 pc} of the LINER/\mbox{Seyfert 1} galaxy NGC\,1097. Using near-infrared high resolution images obtained with 
the ESO-VLT, they reported and argued that these spirals trace cool dust streams, and that they could be the channels by which cold gas and dust flow to a nuclear 
supermassive black hole. Recent works, mostly in the near-infrared,  have analysed several nearby galaxies and revealed 
the presence of different nuclear morphological structures (disk-like, bar-like, spiral-like) responsible for transferring the material from kpc scales down to tens of pc scales 
\citep{Riffel2008, Storchi-Bergmann2010, Riffel2011, Riffel2013, Schnorr-Muller2014}.

The Virgo spiral galaxy NGC\,4501 (M\,88), a Seyfert 2 galaxy, is at a distance of 16.1 Mpc \citep{Ferrarese1996}. This galaxy, an SA(rs)b Hubble type \citep{deVaucouleurs1991},  
was studied by \citet{Silchenko1999} using the Multi-pupil Field Spectrograph attached to the 6-m telescope of the Special Astrophysical Observatory, who found a 
chemically decoupled nucleus within a radius less than 234 pc. \citet{Onodera2004}, using the Nobeyama Millimeter Array, presented a CO (1-0) 2D kinematic analysis of the central 
\mbox{5 kpc} region of NGC\,4501. The main finding was a central gas concentration with a radius of 390 pc, formed by two double peaks separated by \mbox{370 pc}. These authors 
interpreted this gas concentration as originating from gas transfer to the central region by a density wave perturbation due to the spiral arms. 
\citet{Mazzalay2014} accomplished a 2D gaseous and stellar kinematic study of the central $240\,\times\,240$ pc$^2$ of NGC\,4501 with SINFONI/VLT in the H$_2$ 2.12 $\mu$m 
line and detected a blueshifted chain of gas that was interpreted as an outflow in the North-West direction beginning near the centre of NGC\,4501 up to 120 pc. 

NGC\,4501 is undergoing early stage ram pressure stripping, as suggested by some authors \citep{Kenney2004, Crowl2005, Vollmer2007, Vollmer2008, Vollmer2009, Vollmer2012}, and shows 
three to four times less HI than expected for a ``normal" galaxy. This galaxy seems to be in a pre-peak phase of ram pressure stripping \citep{Vollmer2008}. An X-ray luminosity of 
\mbox{$L_X=3.5\times10^{38}$ erg s$^{-1}$} \citep{Li2013} and a bolometric luminosity of \mbox{$L_{Bol}=4.17\times10^{39}$ erg s$^{-1}$} \citep{Wu2013} are reported for NGC\,4501 that is 
close enough to have its central hundreds of parsecs covered by an Integral Field Unit (IFU) instrument, in order to map and verify the existence of streaming motions in the nuclear area. 

We present two-dimensional maps of the gas kinematics and structure within the inner $\sim 500\,\times\,421$ pc$^2$ of NGC\,4501. This work is based on observational 
data acquired through the Gemini Science Archive, gathered with Gemini Multi Object Spectrograph and its Integral Field Unit (GMOS-IFU) \citep{Allington-Smith2002}.

\section{Observations and data reduction}\label{sec:s2}

\begin{figure*}
\hspace*{-1.8cm}
\includegraphics[width=20.5cm, height=8.5cm, keepaspectratio=true]{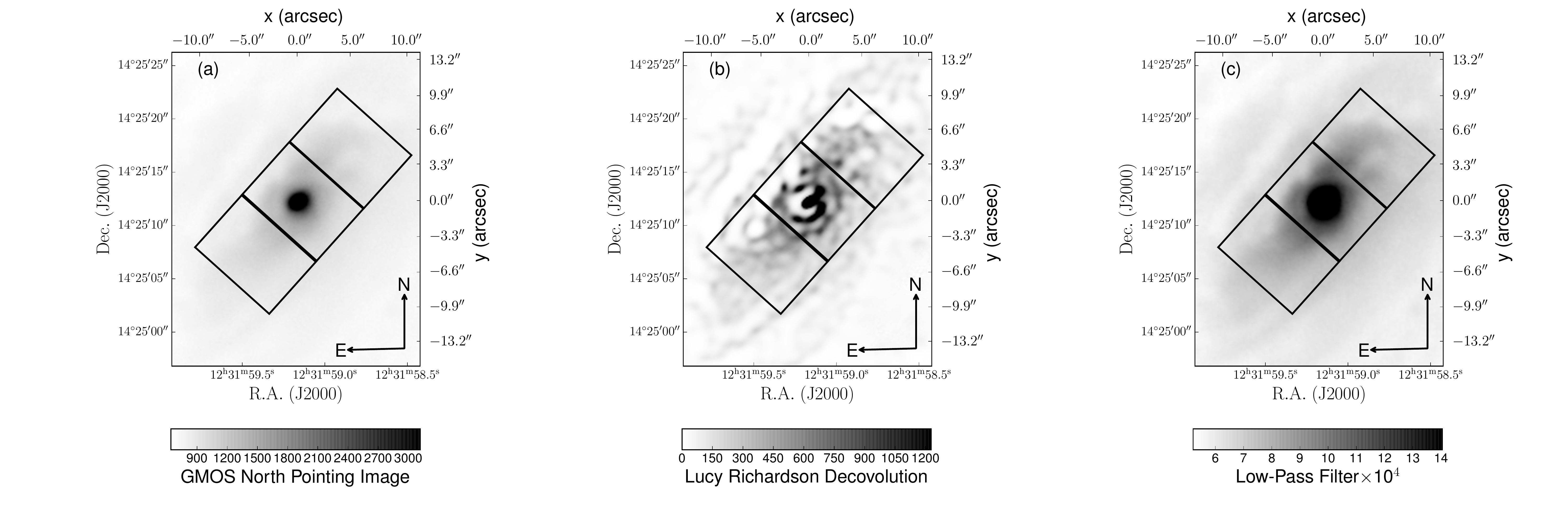}
\caption{(a): Original optical 10-min exposure GMOS-N image in the r-G0303 filter, P.A.=140\degr . 
(b): Lucy-Richardson deconvolution applied to the same image. 
(c): Low-Pass filtered image. 
The three IFU fields are outlined by black boxes. The colorbar units are counts.}
\label{fig1}
\end{figure*}

We used data from the Gemini Science Archive, obtained with the GMOS-IFU at Gemini North telescope, using the R400-G5305 grating and $r$-$G0303$ filter. 
The three IFU fields in the GMOS pointing image, the Lucy Richardson Deconvolution, and a Low-Pass filtered image are displayed in 
Figure~\ref{fig1}. The position angle of the IFU fields is 140\degr, coincident with the photometric position angle of the major axis of NGC\,4501. The central IFU field
is centred at $\alpha=$12h31m59.22s and $\delta=$+14d25m12.69s (J2000), corresponding to the photometric centre of this galaxy. The GMOS-IFU consists of 1000 
science-object fibres (each one of diameter 0.2\arcsec) arranged in a rectangular array of size $7\arcsec\,\times\,5\arcsec$. The spectral setup gave a 
wavelength coverage of 5\,600\,-\,7\,000 \AA\, at a spectral resolution of R$\,\approx\,$3\,500 (85 km s$^{-1}$). Our IFU fields correspond to a rectangular array 
of $\sim\,7\arcsec\, \times\, 15\arcsec$. The spatial resolution of the data is $0.77\arcsec$ \mbox{($\sim$ 60 pc)}. 

\begin{figure*}
\includegraphics[width=\textwidth]{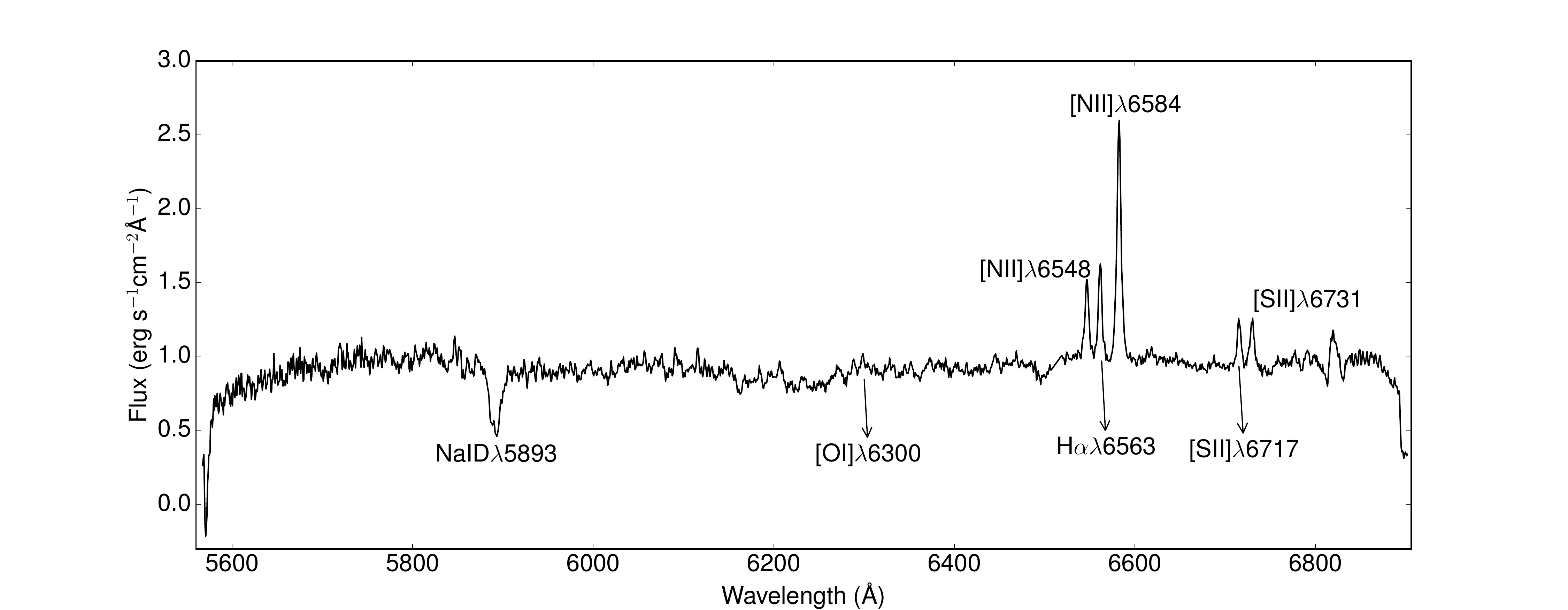}
\caption{The nuclear spectrum of NGC4501, displayed in units of 
$10^{-16}$ erg sec$^{-1}$cm$^{-2}$\AA$^{-1}$, with the emission and absorption lines marked. 
The kinematic maps were derived for the strongest emission lines 
\hbox{[N\,II]$\lambda6584$}, H$\alpha$, and also for the absorption Sodium line.}
\label{fig2}
\end{figure*}

The standard Gemini-IRAF routines were used to carry out bias subtraction, flat-fielding, sky subtraction, and cosmic ray correction. 
The spectra were wavelength calibrated with an accuracy of \mbox{$\leqslant$ 0.3 \AA} and flux-calibrated using the spectroscopic standard star Feige 66, a tertiary standard 
from \citet{Baldwin1984}, as revised by \citet{Hamuy1992}. Finally, the produced data cubes were combined into one reduced cube with dimensions of $101\,\times\,213$ px$^2$ 
(about $\sim 7.34\arcsec\,\times\, 15.5\arcsec$ and $\sim 572\, \times\,1209$ pc$^2$). 
The field of view (FOV) of the analysed data were restricted in order to  have data with a signal to noise ratio (S/N) of about $3-7$ for the emission lines with the strongest 
signal (i.e. \hbox{[N\,II]$\lambda6584$} and H$\alpha$), in the range of \mbox{$88\,\times\,74$ px$^2$} (about $\sim 6.4\arcsec\,\times\,5.4\arcsec$ and $\sim\,500\,\times\,421$ pc$^2$). 
In the case of the NaID$\lambda5892$ absorption line, we fixed the boundaries of our analysed FOV to the interval $\sim\,41\,\times\,47$ px$^2$ (about $\sim3.0\arcsec\,\times\,
3.4\arcsec$ and $\sim 234\,\times\,265$ pc$^2$), to obtain S/N in the same range. 

Figure~\ref{fig2} shows one integrated spectrum in the central $\sim 0.6\,\times\,1.2$ kpc$^2$ of NGC\,4501. It reveals the presence of \hbox{[O\,I]$\lambda6300$}, H$\alpha$, 
\hbox{[N\,II]$\lambda\lambda6548,6584$}, \hbox{[S\,II]$\lambda\lambda6717,6731$} emission lines and the NaID$\lambda5892$ absorption line.

\section{Results}\label{sec:s3}

\subsection{Kinematic analysis: synopsis and methodology}\label{sec:s4}

We constructed the continuum flux, monochromatic flux, velocity field, Full Width at Half Maximum (FWHM), and equivalent width maps for all the lines of the individual spectra, 
with the exception of \hbox{[O\,I]$\lambda6300$} whose S/N ratio is less than 2. The centroid velocities, the fluxes, FWHM and the equivalent width of the observed lines were 
obtained by fitting single Lorentzian profiles to each single line, through the IRAF\footnote{IRAF is distributed by the National Optical Astronomy Observatory, 
which is operated by the Association of Universities for Research in Astronomy (AURA) under cooperative agreement with the National Science Foundation.} task {\it fitprofs}. Tailored 
PyRAF-based scripts were used to perform the fitting task and to produce the 2D maps for each line. The Lorentzian profiles fitted the actual line peaks and wings 
better than the Gaussian profiles, consequently we adopted Lorentzian profiles to derive from the data radial velocities, FWHMs, equivalent widths and fluxes.

\begin{figure*}
\includegraphics[width=\textwidth]{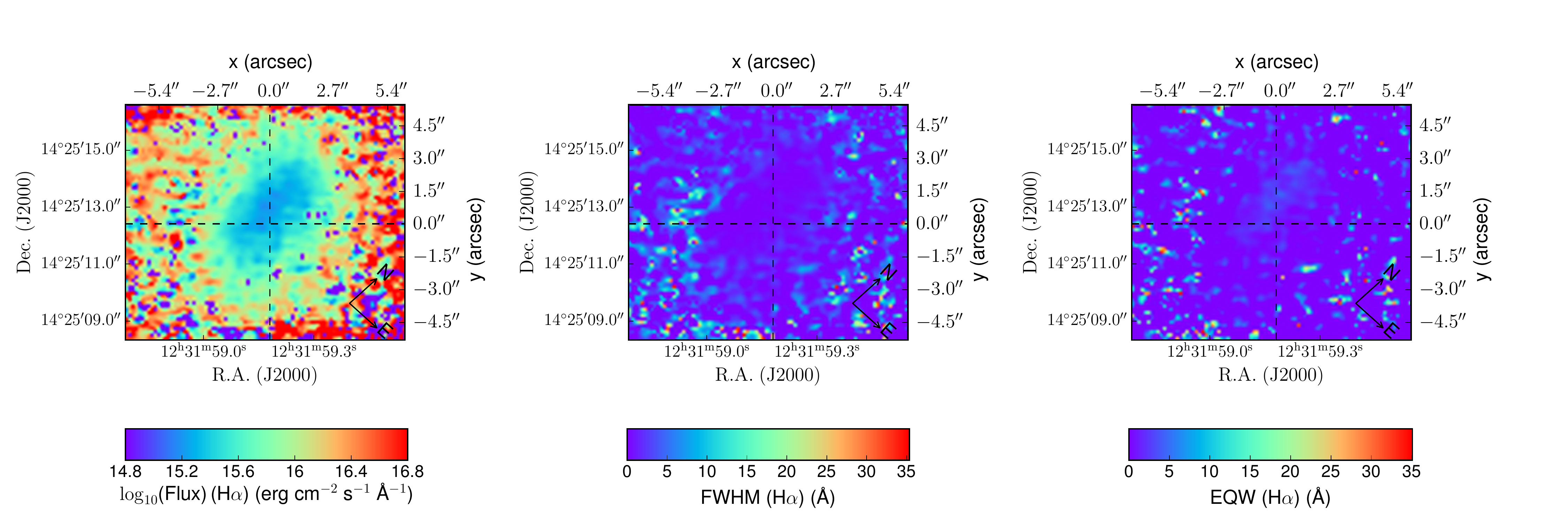}
\caption{Left panel: logarithm of the observed monochromatic flux of \hbox{[N\,II]$\lambda6584$}. 
Middle panel: Lorentzian FWHM of \hbox{[N\,II]$\lambda6584$}. Right panel: equivalent 
width of \hbox{[N\,II]$\lambda6584$}.}
\label{fig3} 
\end{figure*}

\begin{figure*}
\includegraphics[width=\textwidth]{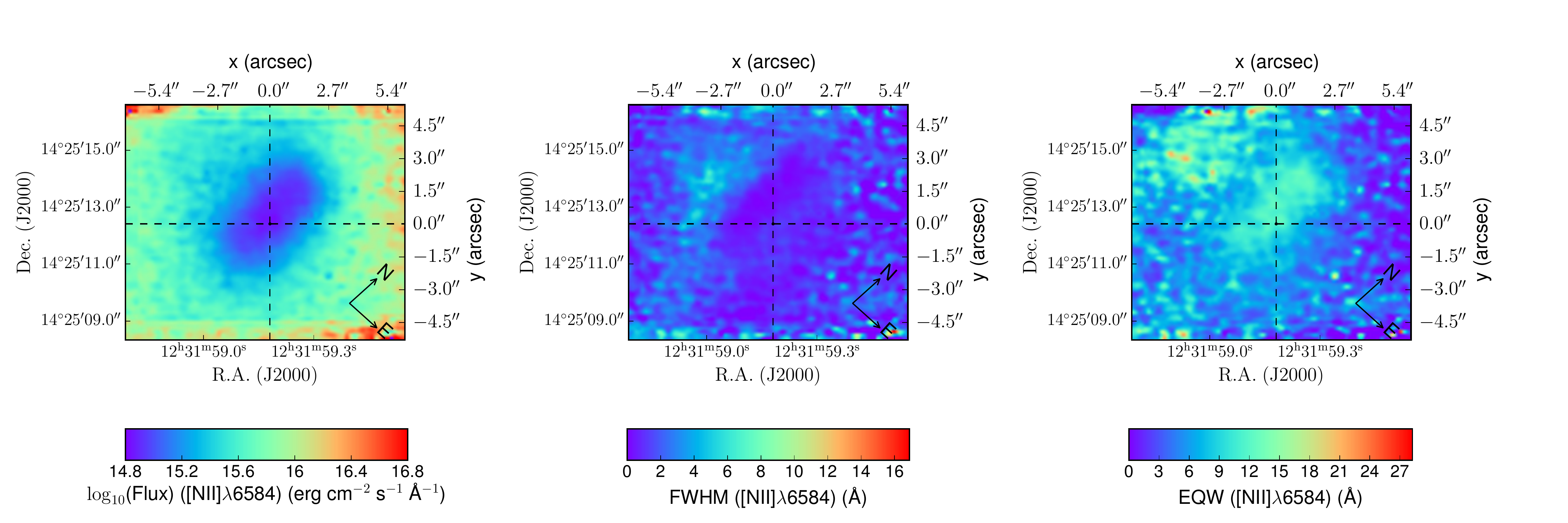}
\caption{Left panel: logarithm of the observed monochromatic flux of H$\alpha$. 
Middle panel: Lorentzian FWHM of H$\alpha$.  Right panel: equivalent 
width of H$\alpha$.}
\label{fig4} 
\end{figure*}

We obtained full spatial maps only for \hbox{[N\,II]$\lambda6584$} and H$\alpha$, due to the more extended and stronger emission with respect to all the remaining lines. 
In the case of NaID$\lambda5892$, we constructed smaller maps for the spatial points with the strongest absorption. The maps of \hbox{[N\,II]$\lambda6548$}  and 
\hbox{[S\,II]$\lambda\lambda6717,6731$}, contain very few points with strong emission, and the kinematics is complex enough to prevent a straightforward interpretation; 
therefore we do not show any maps for these three lines. The line fluxes, FWHM, and equivalent width maps of \hbox{[N\,II]$\lambda6584$} and H$\alpha$ are shown in 
Figures~\ref{fig3} and ~\ref{fig4}.

We corrected the FWHM maps for instrumental resolution, whose value corresponds to $\sim 0.31\arcsec$ ($\sim 4.26$ pxls). The errors were estimated by means 
of Monte Carlo simulations using Poisson statistic model for the data (as already implemented in the {\it fitprofs} task). The result was that the measurement 
errors are less than 5$\%$ in the case of the continuum and line flux and less than 5 km s$^{-1}$ for FWHM, equivalent width and radial velocity. 
We do not show the continuum fluxes, given that the information presented in these maps are very similar to that displayed in the first panels of Figures~\ref{fig3} 
and ~\ref{fig4}. Following, we detail the analysis of the 2D observed velocity fields of \hbox{[N\,II]$\lambda6584$} and H$\alpha$. 

\subsubsection{Kinematics of \hbox{[N\,II]$\lambda6584$} and H$\alpha$}\label{sec:s5}

In an attempt to model the observed velocity fields (OVFs) of \hbox{[N\,II]$\lambda6584$} and H$\alpha$ we applied 
the 2D Locally Weighted Multivariate Regression (LOESS) algorithm of \citet{Cleveland1988} through the python implementation of \citet{Cappellari2013}
to remove noisy regions that could affect the determination of the kinematic position angle (KPA). The KPA of
the OVFs of \hbox{[N\,II]$\lambda6584$} and H$\alpha$ is $114^{\circ}$, however the KPA of the filtered LOESS velocity fields (LVFs) of both lines is 
$137^{\circ}$. The LVFs exhibit relevant non-circular motions, which is the main motivation to produce synthetic velocity fields (SVFs) for
\hbox{[N\,II]$\lambda6584$} and H$\alpha$ to recover gas rotation and expansion velocities. This sort of 2D phenomenological modelling should give us more
insights on the azimuthal, radial and vertical gravitational forces that rule the gas kinematic in the inner zones of NGC\,4501. The analysis of the LVFs was performed 
considering the relation first reported by \citet{Rogstad1971}:
\begin{equation}\label{eqn:e1}
V_{obs}(x,y)=V_{sys}+\left[V_{Rot}(R)\cos{\theta}+V_{exp}(R)\sin{\theta}\right]\sin{i}
\end{equation}

\noindent where $V_{obs}$ is the observed radial velocity at sky coordinates $x$ and $y$, $V_{sys}$ is the systemic velocity, $V_{Rot}$ is the rotation velocity, $V_{exp}$ is the 
expansion velocity, $R$ is the radius in the plane of the galaxy, $i$ is the inclination (INCL) of the plane of the galaxy with respect to the line of sight and $\theta$ is the 
azimuthal angle in the plane of the galaxy given as a function of INCL and KPA. 

\begin{figure*}
\includegraphics[width=\textwidth]{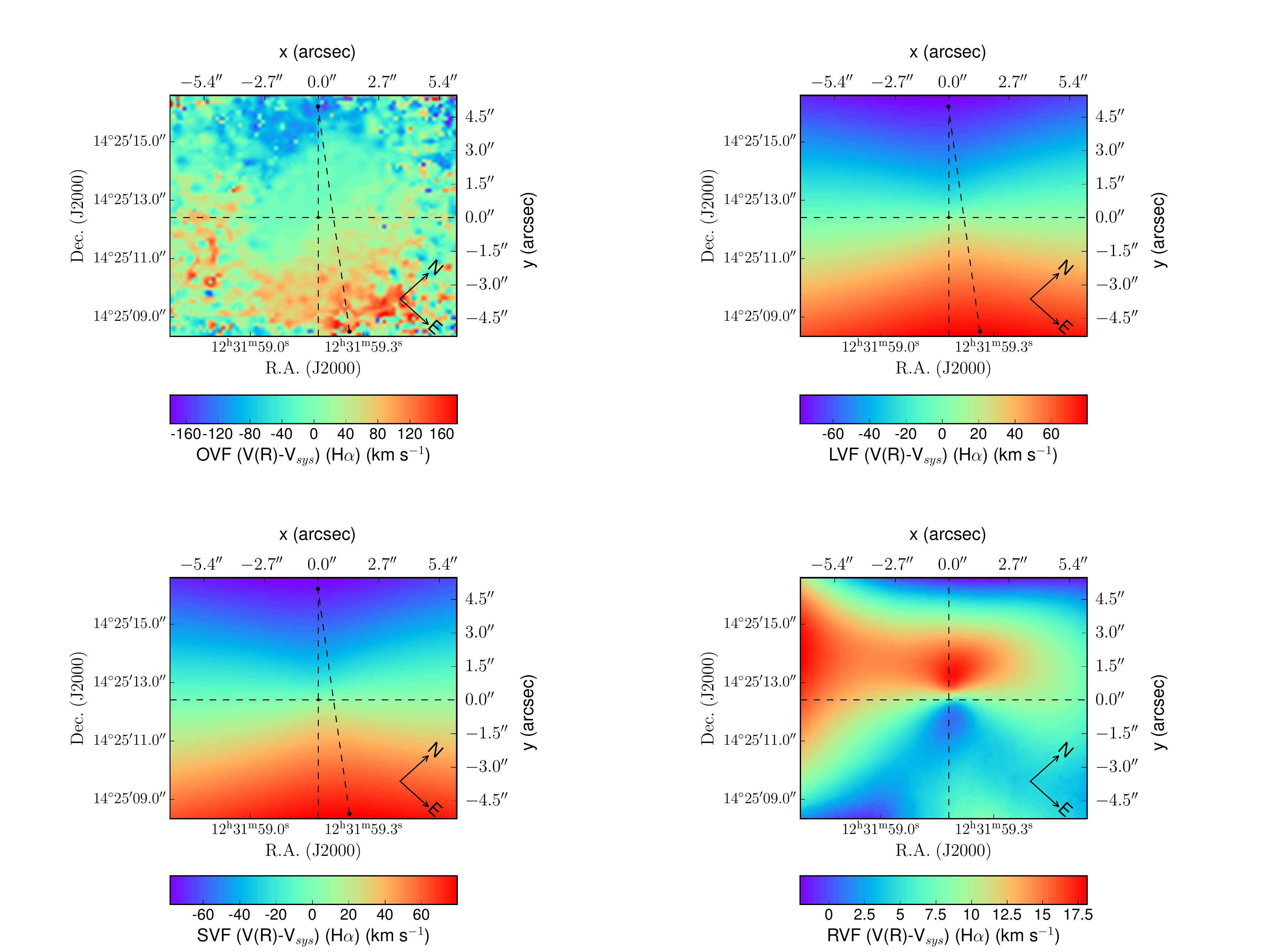}
\caption{Top left: observed velocity field of \hbox{[N\,II]$\lambda6584$}. 
Top right: LOESS filtered velocity field of \hbox{[N\,II]$\lambda6584$}. 
Bottom left: synthetic velocity field of \hbox{[N\,II]$\lambda6584$}.
Bottom right: residual velocity field of \hbox{[N\,II]$\lambda6584$}.
The dashed line connects the two points with maximum velocity gradients denoting 
the kinematic position angle of NGC\,4501.}
\label{fig5} 
\end{figure*}

\begin{figure*}
\includegraphics[width=\textwidth]{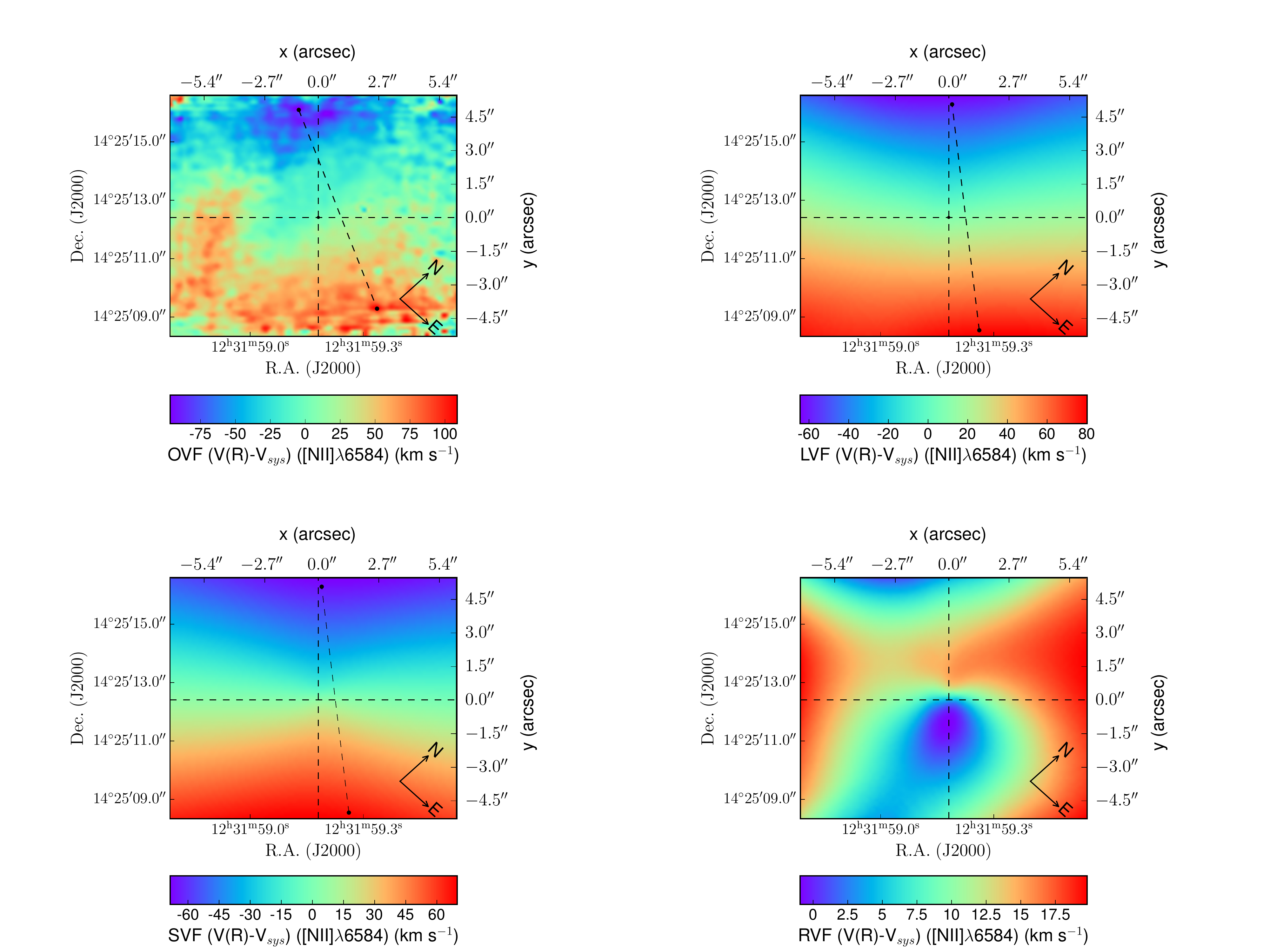}
\caption{Top left: observed velocity field of H$\alpha$. 
Top right: LOESS filtered velocity field of H$\alpha$. 
Bottom left: synthetic velocity field of H$\alpha$.
Bottom right: residual velocity field of H$\alpha$.
The dashed line connects the two points with maximum velocity gradients denoting
the kinematic position angle of NGC\,4501.}
\label{fig6} 
\end{figure*}

We assumed an exponential disk \citep{Freeman1970} as a first guess for the rotation velocity 
and we deduced an expansion velocity by a trial-and-error iterative fitting method aimed to minimize the residual velocity fields (RVFs), obtained subtracting the SVFs to the LVFs. 
In the fitting process the galaxy photometric centre ($\alpha=$12h31m59.22s and $\delta=$+14d25m12.69s (J2000)) and the systemic velocity \mbox{($V_{sys}\,=\,2\,214$ km s$^{-1}$)} 
were fixed. The expansion velocity that returned the minimum RVFs was \mbox{$V_{exp}(R)\,=\,R^{\alpha}+\ln{(R)}$} with $\alpha\,=\,0.75\,\pm\,0.005$ and a maximum value of $\sim\,25$ 
km s$^{-1}$. We also examined Plummer \citep{Plummer1911}, Hernquist \citep{Hernquist1990} and Logarithmic \citep{Binney1984} rotation velocities, yet the least RVFs were obtained with 
the exponential disk model with a maximum value of $\sim\,3$ km s$^{-1}$. The general conclusion of the kinematic analysis of \hbox{[N\,II]$\lambda6584$} and H$\alpha$ was that the LVFs 
were better reproduced by the exponential disk model and the expansion velocity given above; the best fitted kinematic parameters are $KPA\,=\,137^{\circ}\,\pm\,3^{\circ}$ and 
$INCL\,=\,61^{\circ}\,\pm\,2^{\circ}$, in accordance with the corresponding photometric parameters. The LVFs and the SVFs are dominated by the expansion velocity component as one can 
realize from the reported maximum of both velocity components. The OVFs, LVFs, SVFs and RVFs are displayed in Figures~\ref{fig5} and ~\ref{fig6}. 

\subsection{Outflow detection in the inner disc of NGC\,4501}\label{sec:s6}

The NaID$\lambda5892$ absorption line has a low ionization potential (5.1 eV), therefore it is expected to originate in those regions of the interstellar medium shielded from 
ultraviolet radiation, where the hydrogen is mostly neutral. It has been used, by some authors, as a tracer of inflow and outflow of the neutral atomic gas phase in the interstellar 
medium \citep{Phillips1993, Heckman2000, Rupke2005, Davis2012, Rupke2015}.

\begin{figure}
\hspace*{-2.1cm}
\includegraphics[width=12.0cm, height=11.0cm, keepaspectratio=true]{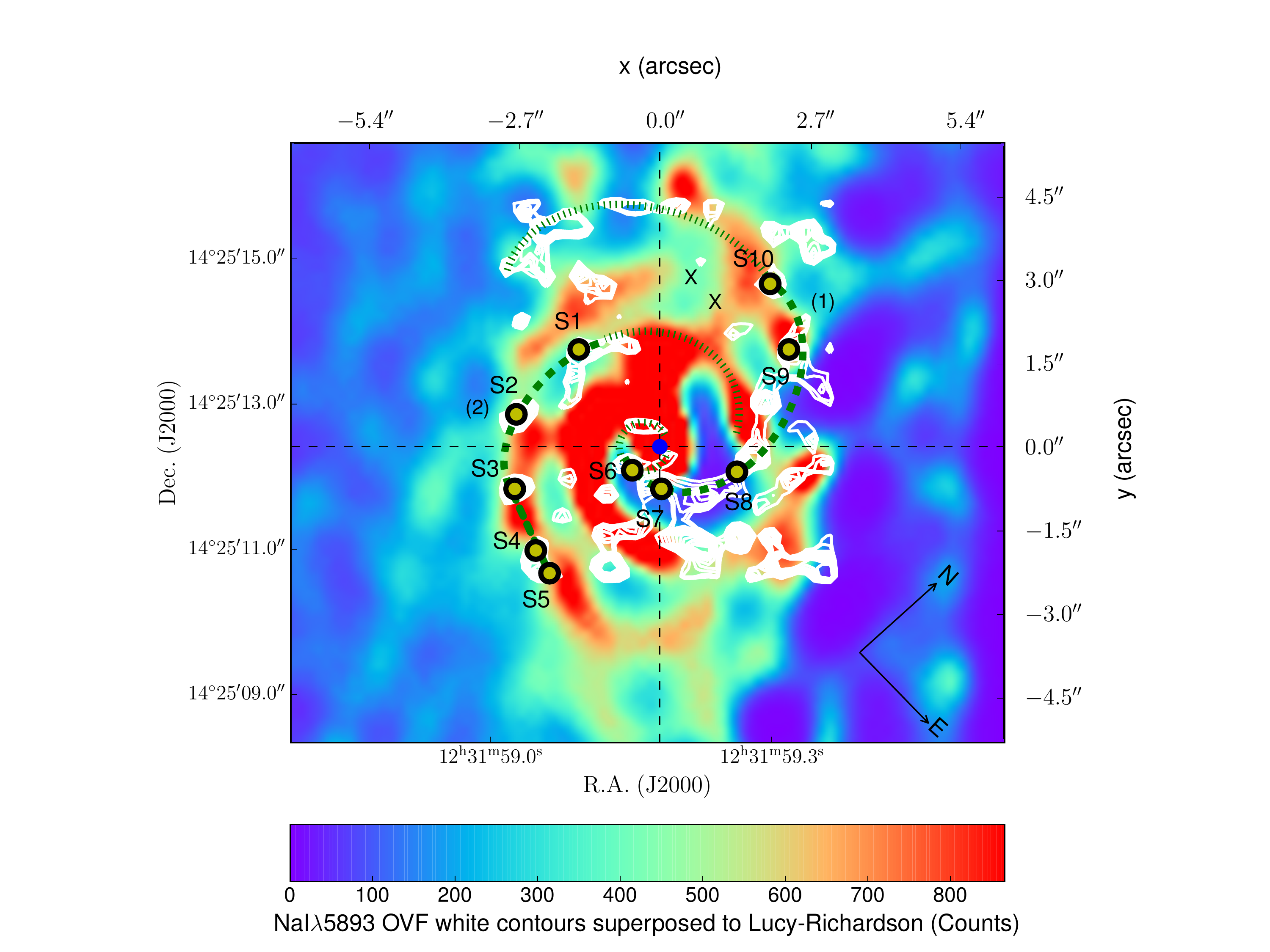}
\caption{Clipped Lucy-Richardson Deconvolution image with superposed white contours of the region 
of NaID$\lambda5892$ OVF which indicates the presence of a possible outflow. The green dashed lines 
suggest possible outflow paths from the central regions of NGC\,4501. The white contour levels of 
the NaID$\lambda5892$ OVF correspond to the radial velocities range from 30 to 102 km s$^{-1}$ 
separated by a step of 8 km s$^{-1}$. The blue dot is the kinematical centre.}
\label{fig7} 
\end{figure}

The NaID$\lambda5892$ velocity field of NGC\,4501 shows filamentary structures that could represent possible gas fuelling towards or from the central regions of this galaxy 
as denoted by the green dashed lines (numbers 1 and 2) in Figure~\ref{fig7}.

Basically, to probe the gas inflow or outflow in a galaxy disc, we need some criteria to establish the actual orientation of this particular galaxy in the sky. The same information
is needed to decide whether a given galaxy has trailing or leading spiral arms. \citet{Sharp1985} (SK85 hereafter), following \citet{deVaucouleurs1958} and \citet{Holmberg1947}, give a 
robust general recipe to ascertain whether a particular spiral galaxy has leading or trailing arms. Their prescription is based on three main pieces of information: (1) 
receding-approaching side, (2) direction of rotation of spiral arms and (3) the tilt of the galaxy (i. e. nearer and farther side to the observer). The same authors showed that the 
interacting spiral galaxy NGC\,5395 has trailing arms, contrary to the claim by \citet{Pasha1982} of a possible candidate with leading arms. \citet{Vaisanen2008} used the SK85 method 
to determine that the luminous infrared galaxy IRAS 18293-3413 has a pair of leading arms. \citet{Repetto2010} employed the same methodology to show that the interacting galaxy pair 
KPG 390 is a trailing pair of spirals.

The two first criteria of SK85 are obvious for NGC\,4501, from the 2D kinematic information collected in this work and from the kinematic of the entire disc of NGC\,4501 
\citep{Chemin2006} (1) and also, for instance, from the SDSS photometry in the g band available in the literature for this galaxy \citep{Adelman2006} (2).
The third criterion of SK85, the tilt of NGC\,4501, could be determined if there were dust lanes around the galactic nucleus and, if this were the case, the nearest side would
be the one which presents the higher difference in intensity than the farthest side, as discussed by SK85. 

\citet{Carollo1998} (CA98 hereafter) analysed Wide Field Planetary Camera 2 (WFPC2) F606W image data of 40 spiral galaxies, including NGC\,4501, providing morphological and 
photometric properties for the entire sample. These authors displayed the $18^{\prime\prime}\,\times\,18^{\prime\prime}$ of the nuclear region of NGC\,4501 
unveiling spiral-like dust lanes surrounding the nucleus.

\begin{figure}
\hspace*{-0.6cm}
\includegraphics[width=11.0cm, height=10.0cm, keepaspectratio=true]{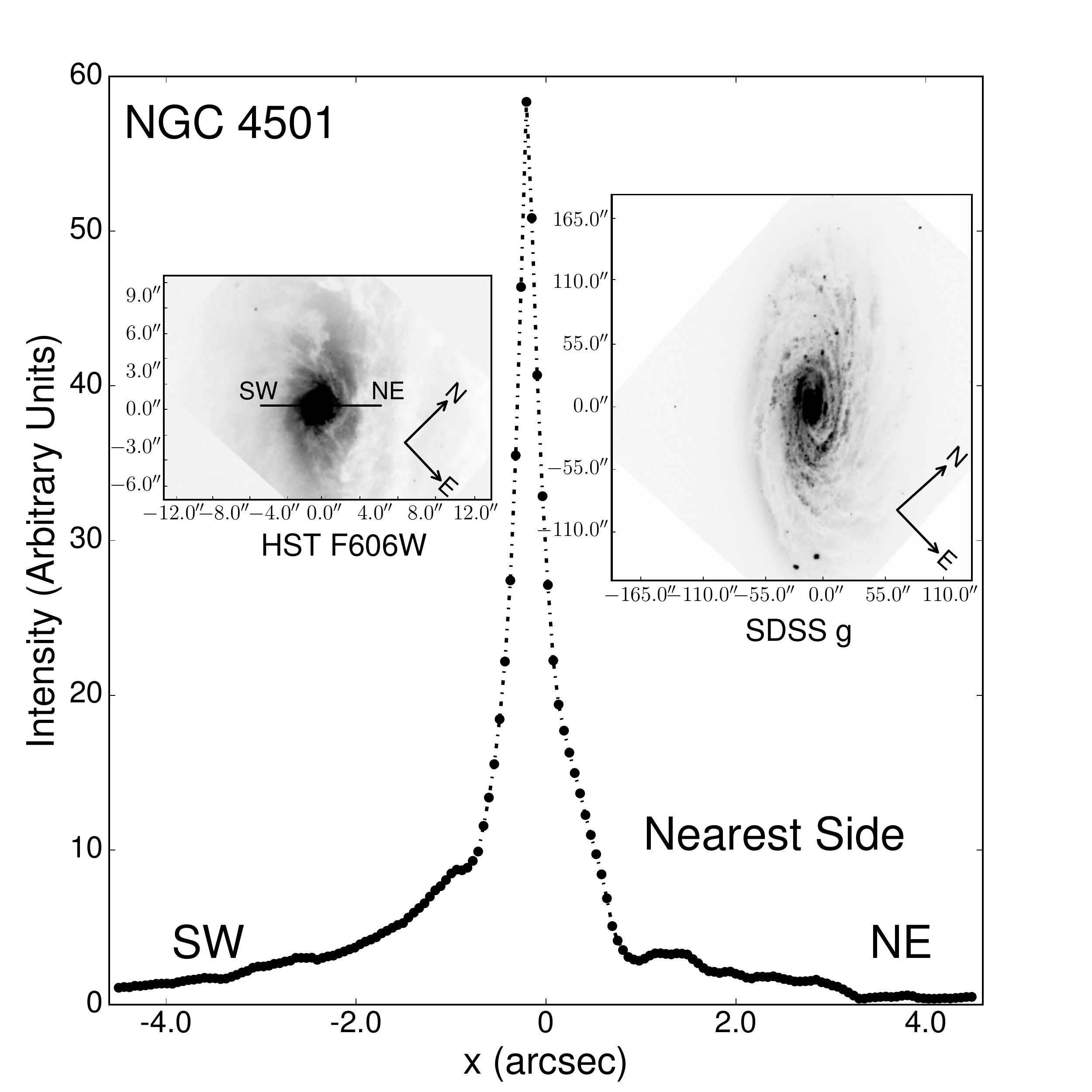}
\caption{Intensity profile along the kinematic minor axis of NGC\,4501. 
Left: HST F606W image of NGC\,4501 from which the intensity profile was derived, 
the NE side is the nearest to the observer because the profile is more abrupt than 
along the SW part. Right: SDSS g band image of NGC\,4501 illustrating the 
counter-clockwise direction pointed by the spiral arms tips of this galaxy.}
\label{fig8} 
\end{figure}

In Figure~\ref{fig8} we show an intensity profile extracted from the WFPC2 F606W image studied by CA98, to determine which side of NGC\,4501 is nearer to the observer. 

The application of the strategy of SK85 to NGC\,4501 produced the following results:

1) The approaching radial velocities are in the northwestern (NW) side of NGC\,4501, whereas the receding radial velocities are in the southeastern (SE) part as suggested by 
Figures~\ref{fig5} and ~\ref{fig6}, therefore the direction of rotation of the disc of NGC 4501 is clockwise. 

2) The tips of the spiral arms of NGC\,4501 point in a counter-clockwise direction as readily noticed from the SDSS g band image showed in Figure~\ref{fig8}.

3) It is evident from Figure~\ref{fig8} that the nearest part of NGC\,4501 to the observer is the NE side, because the profile descends more abruptly than along the SW side.

From these three evidences and Figure~\ref{fig8} we concluded that the filamentary structures S1-S5 and S6-S10 observed in the NaID$\lambda5892$ velocity field of 
Figure~\ref{fig7} correspond to outflowing gas towards the observer and that NGC\,4501 has trailing arms. 

\begin{figure}
\includegraphics[width=\columnwidth]{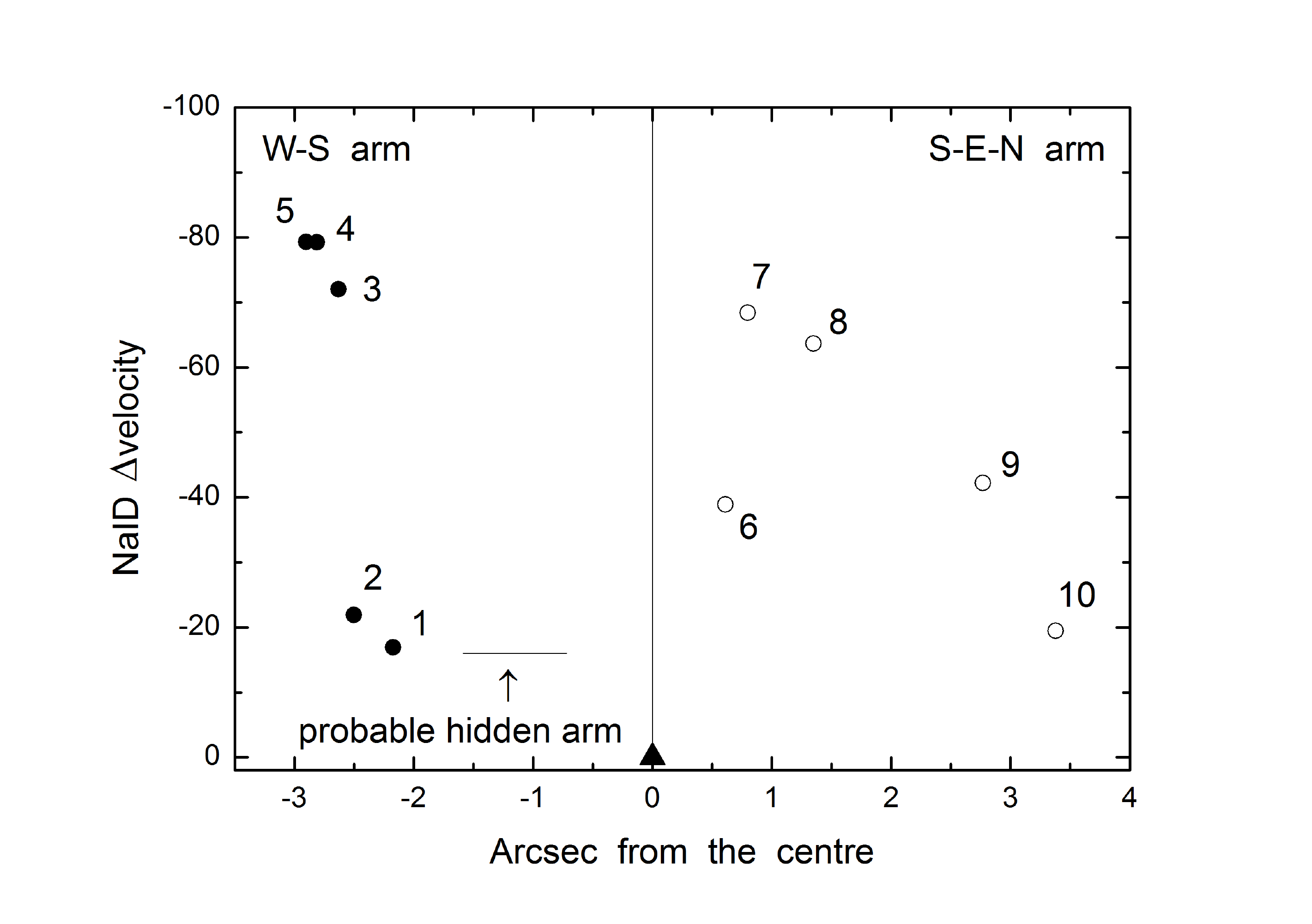}
\caption{The probable outflow arms tracer divided arbitrarily in W-S and S-E-N regions.
 The NaID$\lambda5892$ velocity difference from the central systemic one are displayed. 
 The two likely chains of outflow S1-S5 and S6-S10 regions of Figure~\ref{fig7} are 
 showed (filled and open circles respectively), and a plausible hidden arm section are 
 highlighted. The filled triangle represents the kinematic centre. The N-E 
 direction is the same of Figure~\ref{fig7}.}
\label{fig9} 
\end{figure}

Other empirical proofs of streaming material from the nuclear regions of NGC\,4501 to the observer arise from the analysis of the velocity profiles of the two spiral-like
structures S1-S5 and S6-S10 presented in Figure~\ref{fig9}. The radial velocity of these two pseudo-arms with respect to the central systemic velocity are in the 
range \mbox{10 km s$^{-1} < \Delta$ V $<$ 80 km s$^{-1}$} showing complex gas and dust motion along these inner spirals. The kinematics of \hbox{[N\,II]$\lambda6584$} 
and H$\alpha$ suggests that the gas in the central $500 \times 421$ pc$^2$ of NGC\,4501 is receding to the observer in the S-E and E and approaching to us in the N-W and W. The 
NaID$\lambda5892$ velocity field largely follow the gas kinematics showing clumps and segmented arm-like structures approaching us.\\  
On the other side, \citet{Mazzalay2014} found evidences of outflow from the central regions of NGC\,4501. In spite of the different spatial resolution of their data 
(< 15 pc) and the distinct tracer they used (H$_2$ 2.12 $\mu$m), the black crosses in Figure~\ref{fig7} show a tentative location of 
their outflow detection.

\section{Spectral Synthesis}\label{sec:s7}

The detailed study of the star formation  provides important information on the age distribution along their stellar population components \citep{Krabbe2011}; 
\citep{Faundez-Abans2015}; \citep{Dors2016}. 
We investigate the star formation history of NGC\,4501 using the stellar population synthesis code {\sc STARLIGHT} 
\citep{CidFernandes2004,CidFernandes2005}; \citep{Mateus2006}; \citep{Asari2007}. 

This code is extensively discussed in \citet{CidFernandes2004,CidFernandes2005}, and 
is built upon computational techniques originally developed for empirical population synthesis with 
additional ingredients from evolutionary synthesis models. Briefly, the code fits an observed spectrum with 
a combination of $N_{\star}$ single stellar populations (SSPs) from the Bruzual \& Charlot \citep{Bruzual2003} (BC03) models. 
These models are based on a high-resolution library of observed stellar spectra, which allows for detailed spectral evolution of the 
SSPs across the wavelength range of 3\,200-9\,500 \AA  ~with a wide range of metallicities. We used the Padova's 1994 tracks, as recommended 
by BC03, with the initial mass function of \citet{Chabrier2003} between 0.1 and 100 $M_{\sun}$. 
Extinction is modeled by {\sc STARLIGHT} as due to foreground dust, using the Large Magellanic Cloud average reddening
law of \citet{Gordon2003} with  R$_V$= 3.1, and parametrized by the V-band extinction  A$_V$. 
The SSPs used in this work cover 15 ages, t = 0.001\,, 0.003\,, 0.005\,, 0.01\,, 0.025\,, 0.04\,, 0.1\,,  0.3\,, 0.6\,, 0.9\,, 1.4\,, 2.5\,, 5\,, 11\,, and 13 Gyr, and 
three metallicities, Z = 0.2 Z$_{\sun}$, 1 Z$_{\sun}$, and 2.5 Z$_{\sun}$, adding to  45 SSP components. The fitting is 
carried out using  a simulated annealing plus Metropolis scheme, with regions around emission lines and bad pixels excluded from the analysis.
 
\begin{figure}
\includegraphics[angle=-90,width=\columnwidth]{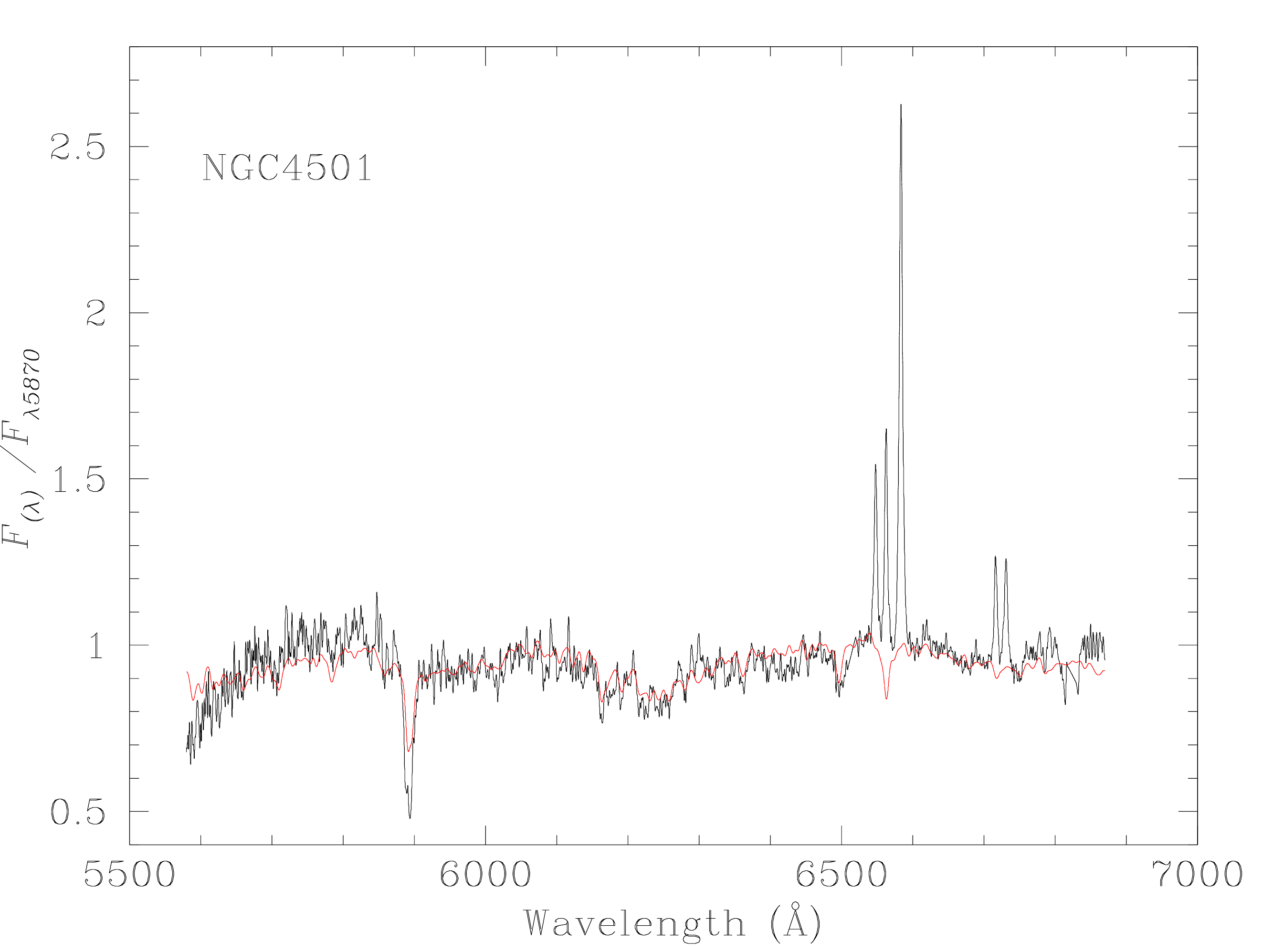}
\caption{Stellar population synthesis for the nuclear region of NGC\,4501. The figure shows the central
bin observed spectrum corrected for reddening (black)
and the synthesized spectrum (red).}
\label{fig10} 
\end{figure}

Figure \ref{fig10}  shows  of the  observed spectrum corrected by reddening and the
model stellar population spectrum for the nuclear region of NGC\,4501. 
Following the prescription of \citet{CidFernandes2005}, we defined a condensed population vector, by binning the stellar populations 
according to the flux contributions into young, $x_{\rm Y}$ ($\rm t \leq 5\times10^{7}$ yr);
intermediate-age,  $x_{\rm I}$ ($ 5\times10^{7} <\rm t \leq 2\times10^{9}$ yr); and 
old, \mbox{$x_{\rm O}$ ( $2\times10^{9} <\rm t \leq 13\times10^{9}$ yr)} components.  The results indicate that 
the central region is dominated by an old stellar population with age of 
$2\times10^{9} <\rm t \leq 13\times10^{9}$ yr. However, a small but non-negligible fraction of young (about 20 \%) and intermediate (about \%10) stars are also present.
The quality of the fitting result is measured by the parameters 
$\chi^{2}$ and  $adev$, whose values are  1.1  and  4.35, respectively. The latter gives the perceptual mean deviation $|O_{\lambda} - M_{\lambda}|/O_{\lambda}$ over all fitted pixels,
where $O_{\lambda}$ and $M_{\lambda}$ are the observed and model spectra, respectively. We found A$_V=0.48$ mag and \mbox{Z= 1  Z$_{\sun}$}.

\section{Diagnostic diagram of the central regions of NGC\, 4501}\label{sec:s8}

To confirm the main ionizing mechanism of the central $500 \times 421$ pc$^2$ of NGC\, 4501, we used the diagnostic diagram introduced by \citet{Lamareille2009} (LM09). 

\begin{figure}
\centering
\includegraphics[width=\columnwidth]{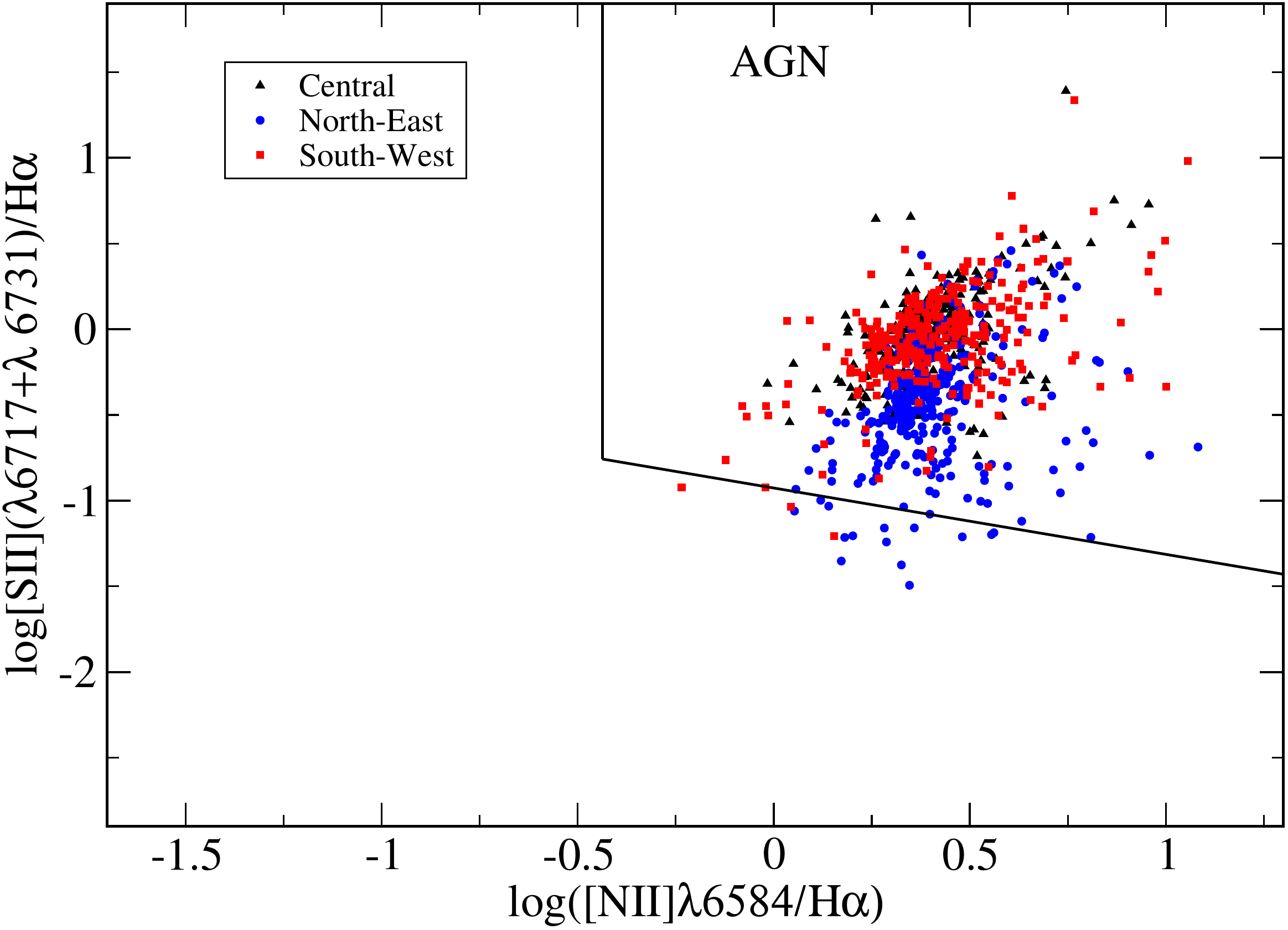}
\caption{Diagnostic diagram log[S II]($\lambda$6717+$\lambda$ 6731)/H$\alpha$ versus log[N II]$\lambda$6584/H$\alpha$ for the nuclear region of NGC\, 4501. The lines 
correspond to the separation between narrow-line AGNs and H II galaxies as proposed by LM09. Black triangles represent measurements at the very centre region, blue circles, 
NE, and red squares, SW regions. The north-east direction is the same of Figure~\ref{fig7}}
\label{fig11} 
\end{figure}

This diagram is an instrument for a preliminary classification of galaxies, based on their \mbox{log[S II]($\lambda$6717+$\lambda$ 6731)/H$\alpha$ and 
log[N II]$\lambda$6584/H$\alpha$} abundances, that separates objects ionized only by massive stars from those containing active nuclei. The diagram 
for the inner parsecs of NGC\, 4501 is displayed in Figure~\ref{fig11}, where the SW, NE, and central zones of NGC 4501 are defined as in 
Figure~\ref{fig7}. 

\citet{Freitas-Lemes2012}, in a study of the Polar Ring galaxy AM 2020-504 and two interacting HII galaxies of the system AM 2229-735, and \citet{Faundez-Abans2009} in the analysis 
of the peculiar ring galaxy HRG 54103, using the LM09 tool, found that the measurements of several regions on the rings were continuously distributed over an 
area which extended upwards from the HII part of the diagram to the AGN one, towards the  locus of the AGN nucleus\rq{} measurement.

The AGN character of NGC\, 4501 is well established. This diagram suggests the existence of some stratification between the central, SW, and NE regions of the central structures of 
this galaxy. In Figure~\ref{fig11}, the top triangles and blue circles are from the nuclear signal approaching us, the rest of the blue circles are from the NE region, 
and the red squares are measurements in the SW region receding from the observer.

\section{Discussion}\label{sec:s9}

The study of the nuclear morphological structures formed through
the transport of material towards the central regions of galaxies represents 
a fundamental topic to understand the complex phenomenon of nuclear activity.
As mentioned in the Introduction, several central structures have been already
identified and analysed from a kinematic and physical point of view, in some 
galaxies, establishing a scenario where nuclear bar-like, disk-like, and 
spiral-like features are playing a pivotal role in the transfer of gas up 
to scales of tens-of-parsecs. At the present time, the number of studied galaxies is 
small, and it is difficult to overcome the tens of parsecs limit. Nevertheless, the 
above-mentioned studies represent a first and fundamental step to achieve a deeper 
understanding of the physics and kinematics that rule the fuelling of the central engine. 

Bearing in mind the framework settled by these previous works 
and with similar spatial resolution, the present analysis attempts to find evidence 
of kinematic structures that could be connected with the streaming motion of 
gas towards the centre of NGC\,4501. We have used GMOS-IFU archival data to perform a detailed study of
the 2D gas flux, radial velocity, FWHM, equivalent width and structures within the 
central $\sim\, 500 \times 421$ pc$^2$ of the Seyfert 2 galaxy NGC\,4501 a member of the 
Virgo cluster of galaxies. We used imaging filtering techniques and 2D kinematic analysis
applied respectively to the GMOS pointing image and IFU data cube to investigate the streaming 
gas motion in the inner zones of NGC\,4501. 

Additionally we provide evidences of the prevailing AGN behaviour of NGC\,4501 and the dominant old stellar
populations residing in the interior disc of this galaxy separately through diagnostic diagram techniques 
and spectral synthesis analysis. The result of the spectral synthesis study confirms in part the AGN 
preponderance with respect to starburst in the inner hundreds of parsecs of NGC\,4501.

The 2D kinematic of \hbox{[N\,II]$\lambda6584$} and H$\alpha$ gives some information about the amplitude of 
the stellar gravitational potential and the decoupling between stars rotation and gas diffusive motion in the 
inward regions of NGC\,4501. A fundamental result of the 2D kinematic analysis is that gas diffusion dominates 
over stellar rotation in the central zones of NGC\,4501 because of the decrease of the total stellar mass at 
smaller radii.

The plausible detection of outflowing material across the inner spiral arms S1-S5 and S6-S10, confirms the 
predominance of non circular motions and intricate gas and dust dynamics on different levels in the inner 
zones of NGC\,4501.

In general galaxies are 3D aggregates of several gravitating constituents, principally stars, gas
and dark matter. In the central regions of galaxies the gravitational contribution of dark matter can be neglected 
with respect to that of baryons according, for instance, to the analysis of \citet{Linden2014}. The author compares 
the masses of baryons and dark matter in the inner 100 pc of the Galaxy using the model of \citet{McMillan2011} showing 
the gravitational predominance of baryons over dark matter, based on these results we decided to consider exclusively the 
stellar and gas gravitational components.

From a kinematic point of view, the 2D deprojection of the observed radial velocities
of each point of an actual 3D galaxy originates a measure of the kinematic of all the gravitational 
components present in the analysed galaxy. The rotation and expansion velocity components extracted from a 2D velocity 
field generated by a 3D observed data cube of a given galaxy are valuable information about the dominant motions 
within this particular galaxy disc and allow, in principle, the separation of the kinematics of the stellar and gas components 
assuming prevalent circular movements associated to the stellar gravitational potential. These considerations are applicable 
to a real galaxy as far as the circular motions of the stars predominate over the radial and vertical streaming motions. The latter 
approximation could not completely hold in the central region of galaxies due to the increase of stellar random motions and the 
corresponding decrease of stellar ordered rotation.

In the Galaxy, some studies (see \citet{Genzel2010} for a complete set of references) depicted the composite stellar kinematics of the 
central regions with much more detail and accuracy than for external galaxies, therefore, we shortly recapitulate those findings 
about the stellar motions in the inner zones of our galaxy in order to comment briefly about the results of the 2D gas kinematic analysis 
of NGC\,4501.   

The very central regions of our galaxy ($<$ 10 pc), as determined by stellar proper motion measurements, are mostly characterised 
by an overall rotation of old and young stars in a nuclear cluster and in a disc with different inclinations and 
position angles with respect to the sky plane respectively. Another component of massive O and Wolf-Rayet stars display azimuthal motion in 
the opposite direction to the disc of young stars mentioned above \citep{Trippe2008, Schodel2009, Paumard2006, Lu2009}.

The actual stellar kinematical picture of the Galaxy could be much more complex and additional non circular motions could originate in the 
intersection between azimuthal motions on surfaces with different orientation with respect to the sky plane as indicated by the researches 
quoted above and also by other observational and theoretical studies \citep{Bartko2009, Kocsis2011}. 

The kinematic analysis performed in the present work utilizes two emission 
lines (\hbox{[N\,II]$\lambda6584$} and H$\alpha$) as gaseous tracers of the stellar gravitational potential of NGC\,4501. Albeit we are aware that 
our knowledge of the stellar motion is indirect, the gaseous tracers could be influenced by physical forces other than gravity (e.g. radiative 
and magnetic force) and the spatial resolution of our data is good ($\sim$ 60 pc), but not so high as that of the quoted investigations about the Galaxy, 
in the particular case of NGC\,4501 we successfully separate rotation from non circular motions and in conformity with previous works quoted in the Introduction 
we argue that, to a first approximation, the circular velocities detected in the inward zones of NGC\,4501 primarily represent the remnant of the radial variation 
of the stellar gravitational potential, whereas the non circular velocities are more likely associated to gas radial and vertical (deprojected) motions. These 
findings, together with the other results derived in the present study of NGC\,4501 constitute a suitable approach to disentangle the kinematic and physical conditions 
in the central disc of this galaxy. In particular, a future 3D kinematic analysis of NGC\,4501 should take into account in a more realistic way the actual three 
dimensional kinematics of stars and gas and their interaction on different vertical levels of the disc of NGC\,4501.  

\section{Conclusions}\label{sec:s10}

In this article we performed a 2D kinematic study of the inner regions of NGC\,4501 in an attempt to ascertain the azimuthal and radial gas velocities
in the internal disc of NGC\,4501. We accomplished a spectral synthesis and diagnostic diagram analysis in an effort to deduce the true nature of the 
central engine of NGC\,4501. We collected empirical evidences of outflowing material from the central parsecs of NGC\,4501. 

In summary our findings about the kinematical and physical properties of the inner $500 \times 421$ pc$^2$ of NGC\,4501 are the following:

\begin{itemize}
 \item The central regions of NGC\,4501 are dominated by expansion velocity probably associated to dissipative gas processes. 
 \item Old stars predominate in the central zones of NGC\,4501 suggesting weak or absent star formation activity. 
 \item The central part of NGC\,4501 exhibits a predominant AGN character.
 \item We detected two inner pseudo-spirals that likely represent chains of outflowing material from the centre of NGC\,4501. 
\end{itemize}

This work presents a combination of different analysis techniques to gather a broader knowledge about the kinematical and physical properties
of the inner regions of \mbox{NGC 4501}. The itemized results elucidate some global aspects of the phenomenological behaviour of gas and stars in the central
parsecs of NGC 4501.

\section*{Acknowledgments}

R. P. thanks LNA for postdoctoral fellowship through CNPq grant 313998/2013-2.\\
I. R. thanks the support of FAPESP agency, process 2013/17247-9.\\
P.F-L thanks the support of CAPES.\\
This work was partially supported by the Brazilian Minist\'{e}rio da Ci\^{e}ncia, Tecnologia e Inova\c{c}\~{a}o (MCTI), Laborat\'{o}rio Nacional de 
Astrof\'{i}sica (MCTI/LNA). Based on observations obtained at the Gemini Observatory (acquired through the Gemini Science Archive and processed 
using the Gemini IRAF package), which is operated by the Association of Universities for Research in Astronomy, Inc., under a cooperative agreement with 
the NSF on behalf of the Gemini partnership: the National Science Foundation (United States), the National Research Council (Canada), CONICYT (Chile), 
Ministerio de Ciencia, Tecnolog\'{i}a e Innovaci\'{o}n Productiva (Argentina), and Minist\'{e}rio da Ci\^{e}ncia, 
Tecnologia e Inova\c{c}\~{a}o (Brazil). This research made use of the NASA/IPAC Extragalactic Database (NED), which is operated by the Jet Propulsion Laboratory,
California Institute of Technology, under contract with the National Aeronautics and Space Administration. This research has made use of the NASA/IPAC Infrared
Science Archive, which is operated by the Jet Propulsion Laboratory, California Institute of Technology, under contract with the National Aeronautics and Space
Administration. Based on observations made with the NASA/ESA Hubble Space Telescope, and obtained from the Hubble Legacy Archive, which is a collaboration between
the Space Telescope Science Institute (STScI/NASA), the Space Telescope European Coordinating Facility (ST-ECF/ESA) and the Canadian Astronomy Data Centre 
(CADC/NRC/CSA).We acknowledge the use of the Kapteyn package \citep{KapteynPackage} and the ROOT package \citep{ReneBrun}. We used the CAPLOESS2D routine of 
\citet{Cappellari2013}, which implements the 2D multivariate LOESS algorithm of \citet{Cleveland1988}.


\label{lastpage}
\end{document}